\documentclass[aps,prd,showpacs,nofootinbib,superscriptaddress]{revtex4}
\usepackage{eurosym}
\usepackage{epsfig}
\usepackage{graphicx}
\usepackage{graphics}
\usepackage{color}

\newcommand{\rrv}[1]{{#1}}
\newcommand{\rr}[1]{{#1}}

\begin{document}

\title{Classical limit for Dirac fermions with modified action in the presence of the black hole}
\author{M. Lewkowicz}
\affiliation{Physics Department, Ariel University, Ariel 40700, Israel}
\author{M.A. Zubkov \thanks{%
On leave of absence from Institute for Theoretical and Experimental Physics,
B. Cheremushkinskaya 25, Moscow, 117259, Russia}}
\affiliation{Physics Department, Ariel University, Ariel 40700, Israel}
\email{zubkov@itep.ru}

\begin{abstract}
We consider the model of Dirac fermions coupled to gravity as proposed in
\cite{VolovikBH},
in which superluminal velocities of
particles are admitted. In this model an extra term is added to the
conventional Hamiltonian that originates from Planck physics. Due to this
term a closed Fermi surface is formed in equilibrium inside the black hole.
In this paper we propose the covariant formulation of this model and analyse
its classical limit. We consider the dynamics of gravitational collapse. It
appears that the Einstein equations admit a solution identical to that of
the ordinary general relativity. Next, we consider motion of particles in
the presence of the black hole. Numerical solutions of the equations of
motion are found which demonstrate that the particles are able to escape
from the black hole.
\end{abstract}

\date{\today}
\maketitle
\affiliation{Physics Department, Ariel University, Ariel 40700, Israel}
\affiliation{Physics Department, Ariel University, Ariel 40700, Israel}




\section{Introduction}

The Schwarzschild black hole (BH) solution \cite{Schwarzschild} may be
brought to the form, which is especially useful for the consideration of the
motion of matter. This is the so - called Painlev\'{e} - Gullstrand (PG)
black hole \cite{Gullstrand,Painleve}. In the corresponding reference frame
space - time looks like flat space falling down to the center of the BH with
velocity that depends on the distance to the center. Such a representation
also exists for the Reissner-Nordstrom BH and even for the Kerr BH \cite%
{Hamilton:2004au,Doran:1999gb}. The structure of the BH solution in the PG
reference frame prompts to consider the analogy to the motion of fluid in
laboratory. Such an analogy has been considered in the framework of the
theory of $^{3}$He superfluid in \cite{Volovik:1999fc}. It allows to
calculate in a demonstrative way the Hawking radiation \cite{Hawking:1974sw}
(see also, for example, \cite%
{Parikh:1999mf,Akhmedov:2006pg,Jannes:2011qp,Volovik2003} and references
therein).

The consideration of Dirac fermions in the PG reference frame leads to the
unexpected observation, that inside the BH the fermion states with vanishing
energy form the surface in momentum space. In equilibrium, at vanishing
temperature, it becomes the Fermi surface and separates the region of
occupied quantum states from the region of the vacant ones. For the ordinary
Dirac fermions such a surface is open and infinite. The analogy with
condensed matter physics suggests that the particle Hamiltonian is to be
modified in such a way so that the resulting Fermi surface will become
finite and closed. It has been proposed in \cite{VolovikBH}, that such a
modification occurs due to the Planck physics. The corresponding term has
been added to the Dirac Hamiltonian. It leads to several consequences for
the BH physics. First of all, the analogy to the BH in the PG reference
frame has been found within the class of the recently discovered materials
called Weyl semimetals \cite%
{semimetal_effects6,semimetal_effects10,semimetal_effects11,semimetal_effects12,semimetal_effects13,Zyuzin:2012tv,tewary,16}%
. Bloch electrons within those materials behave similarly to the elementary
particles. In the special type of such materials called the type II Weyl
semimetals (WSII) \cite{W2} the dependence of energy of electrons on momenta
\cite{VZ} possesses an analogy to that of the particles in the interior of
the BH \cite{VolovikBHW2,NissinenVolovik2017a}.

In \cite{Z2018,Z2018_2} it was noticed that if a closed finite Fermi surface
inside the BH is formed, there should exist particles that escape from the
BH without tunneling. In the present paper we take a step back and consider
the model proposed in \cite{VolovikBH} on the classical level. We suppose
that a careful consideration of the classical dynamics should precede the
more sophisticated discussion of the quantum BH, although the extra term
added to the particle Hamiltonian becomes relevant at Planck energies.

This extra term contains the time - like four - vector $n_{\mu }$. In the
Painlev\'{e} - Gullstrand reference frame it marks the direction of time. In
the covariant theory there should be no such preferred direction of time. We
assume that it appears as a result of a dynamical symmetry breaking. This
symmetry breaking, in turn, results in the appearance of the massless
Goldstone modes. Here we shall not discuss the physics of those massless
excitations. 

First of all we propose the covariant formulation of the discussed model. It
contains the vector field $n_{\mu }$, which, after the spontaneous
breakdown, acquires its particular value that points in the direction of
time in the Painlev\'{e} - Gullstrand reference frame. The classical
equations of motion for the corresponding point - like particles admit
motion with superluminal velocity. Therefore, unsurprisingly, the particles
may escape from the black hole already on the classical level. For the
discussion of the theories that admit superluminal velocity of particles we
refer to \cite{superluminal}.

The paper is organized as follows. First of all, in Sect. \ref{Section_DF}
we recall the general properties of the Dirac fermions in the PG reference
frame in the presence of the charged BH. In Section \ref{SectCov} we propose
the covariant formulation of the model of \cite{VolovikBH} and derive its
classical limit. In Sect. \ref{SectT} we derive the expression for the
stress - energy tensor of the noninteracting particles in the PG reference
frame. In Sect. \ref{collapse} the gravitational collapse in this model is
considered. In Sect. \ref{SectClass} we present the results of the numerical
solution of the classical equations of motion for the motion of particles in
the presence of the existing black hole (at the stage when the gravitation
collapse is finished). The physical significance of the results is dicussed
in the concluding section \ref{SectConcl}.

\section{Dirac fermions in the black hole in the Painleve - Gullstrand reference frame}

\label{Section_DF}

\begin{figure}[h]
\centering
\includegraphics[width=10cm]{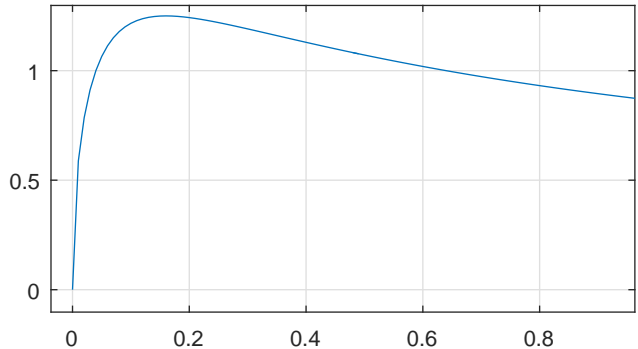}
\caption{Velocity of "vacuum" as a function of $r$ for $Q=0.4, M = 0.5 \,
m_P$.}
\label{fig.v}
\end{figure}


In the present paper we will mainly be interested in neutral black holes.
However, we start our consideration from the ansatz that corresponds to a
charged black hole. The charge is assumed to be small, so that it modifies
the BH solution in the small vicinity of the BH center. We will add also
another modification in the small vicinity of the center, which will cause
the metric to be regular everywhere. Thus we start from the charged black
hole metric in the PG coordinates
\begin{equation}
ds^{2}=dt^{2}-(d\mathbf{r}-\mathbf{v}(\mathbf{r})dt)^{2},  \label{PGQ}
\end{equation}%
Here the expression
\begin{equation}
\mathbf{v}=-\frac{1}{m_P }\frac{\mathbf{r}}{r}\,\sqrt{\frac{2M}{r}-\frac{%
Q^{2}}{r^{2}}}
\end{equation}%
may be considered as a velocity of space falling towards the center of the
BH. $Q$ is the charge of the BH, $m_P $ is \rrv{the Plank mass}, while $M$ the
BH mass.

The vielbein is given by
\begin{equation}
E_{a}^{\mu }=\left(
\begin{array}{cc}
1 & \mathbf{v} \\
0 & 1%
\end{array}%
\right) ,
\end{equation}%
and the inverse vielbein $E_{a}^{\mu }$ is
\begin{equation}
e_{\mu }^{a}=\left(
\begin{array}{cc}
1 & -\mathbf{v} \\
0 & 1%
\end{array}%
\right) .
\end{equation}%
The metric is equal to
\[
g_{\mu \nu }=e_{\mu }^{a}e_{\nu }^{b}\eta _{ab}
\]%
where $\eta _{ab}=\mathrm{diag}\,(1,-1,-1,-1)$.

The action for the Weyl fermions has the form
\begin{eqnarray}
S_{R,L} &=&\int d^{4}x\,\mathrm{det}^{-1}(\mathbf{E})\,\bar{\psi}_{R,L}(x)%
\Big(iE_{0}^{0}\partial _{t}\pm \frac{i}{2}\tau ^{a}\Big[E_{a}^{k}{\partial }%
_{k}+\partial _{k}E_{a}^{k}\Big]\Big)\psi _{R,L}(x)  \nonumber \\
&=&\int d^{4}x\,\bar{\psi}_{R,L}(x)\Big(i\partial _{t}-H^{R,L}(-i\partial )%
\Big)\psi _{R,L}(x),
\end{eqnarray}%
where symbols $R$ ($L$) mark the right - handed $/$ left - handed fermions.
Their Hamiltonians are
\begin{equation}
H^{R,L}(\mathbf{p})=\pm \mathbf{p}\sigma -\mathbf{p}\mathbf{v}
\end{equation}%
The Dirac mass term mixes the right - handed and the left - handed fermions
\begin{eqnarray}
S_{m} &=&-m\sum_{R,L}\int d^{4}x\,\mathrm{det}^{-1}(\mathbf{E})\,\bar{\psi}%
_{R,L}(x)\psi _{L,R}(x)  \nonumber \\
&=&-m\int d^{4}x\,\sum_{R,L}\bar{\psi}_{R,L}(x)\psi _{L,R}(x)
\end{eqnarray}%
It appears, that the spin connection does not enter this action.
Next, following \cite{VolovikBH}, we introduce the term that breaks the
Lorentz invariance:
\[
S_{P}=-\frac{1}{\mu}\sum_{R,L}\int d^{4}x\,\mathrm{det}^{-1}(\mathbf{E})\,\Big[\pm \,\Big(\vec{\partial}\bar{%
\psi}_{R,L}(x)\Big)\Big(\vec{\partial}\psi _{L,R}(x)\Big)\Big]
\]

\rrv{Here parameter $\mu$ is assumed to be of the order of the Plank mass.} For $r<r_{0}=\frac{Q^{2}}{2M}$ the velocity $v$ becomes imaginary. We guess
that due to interaction with matter the dependence of $v$ on $r$ is modified
within the BH, and $v$ remains real and tends to zero at $r=0$ (see also
\cite{Hamilton:2004au}). In the present paper we model this form of $v$ via
the following modification
\begin{equation}
v(r)=\frac{1}{m_P }\sqrt{\frac{2M}{r+\epsilon }-\frac{Q^{2}}{(r+\epsilon
)^{2}}}
\end{equation}%
with
\[
\epsilon =\frac{Q^{2}}{2M}
\]%
the resulting form of $v(r)$ is represented in Fig. \ref{fig.v} in the
system of units with $m_P =1$ at $M=\frac{m_P }{2}$and $Q=0.4$.

The two horizons are placed at
\begin{equation}
r_{+}=\frac{M+\sqrt{M^{2}-Q^{2}m_P^{2}}}{m_P^{2}}-\epsilon
\end{equation}%
and
\begin{equation}
r_{-}=\frac{M-\sqrt{M^{2}-Q^{2}m_P^{2}}}{m_P^{2}}-\epsilon
\end{equation}%
For $r>r_{+}$ there are the ordinary Dirac fermions. At $m=0$ the Fermi
point appears. Between the two horizons $r_{-}<r<r_{+}$ at $m=0$ there is
the type II Dirac point, while $|\mathbf{v}|$ is larger than light velocity.

\begin{figure}[h]
\centering
\includegraphics[width=10cm]{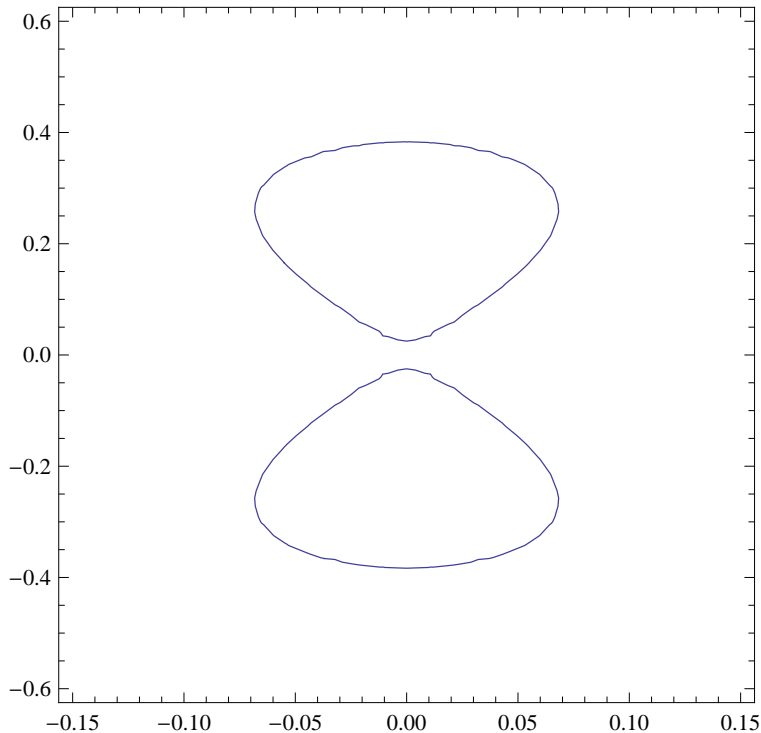}
\caption{The typical Fermi surface form in the plane $(p_r, p_\bot)$ within
the black hole at the values of parameters $M = 0.5 \, m_P$, $m=0.01 \, m_P$%
, $\mu = m_P$, $Q = 0.4$. For these values the external horizon is placed at $r_+ =
0.64/m_P$. We represent the Fermi surface at $r=0.5/m_P$.}
\label{fig.F}
\end{figure}

Neglecting the derivatives of $\mathbf{v}$ we come to the following
expression for the particle energy
\begin{equation}
\mathcal{E}(\mathbf{p})=\pm \sqrt{m^{2}+\mathbf{p}^{2}+\frac{\mathbf{p}^{4}}{%
\mu ^{2}}}-\mathbf{p}\hat{\mathbf{r}}\,v(r)  \label{Energy0}
\end{equation}%
Thus assuming that $\mathbf{v}$ varies slowly, we come to the conclusion
that between the two horizons the particle energy vanishes along the closed
surface in momentum space. Its form is represented in Fig. \ref{fig.F} for
the particular choice of parameters. It is worth mentioning, that the
Hamiltonian of the form of Eq. (\ref{Energy0}) admits motion with the
superluminal velocity on the classical level. Therefore, unsurprisingly, in
the considered model the particles are able to escape from the Black hole.


\section{Covariant formulation of the theory and its classical limit}

\label{SectCov}

Here we propose the covariant modification of the model of \cite{VolovikBH}
considered in the previous section. Namely, we consider the Dirac field $%
\Psi $ with action
\rr{\begin{eqnarray}
S &=&\int d^{4}x\,\mathrm{det}^{-1}(\mathbf{E})\,\Big( \bar{\Psi}(x)\gamma ^{0}%
\Big[i\gamma ^{a}E_{a}^{\mu }D_{\mu } - m \Big]\Psi(x) \nonumber \\
&&+\frac{1}{\mu}\Big[D_\mu\bar{\Psi}(x)\Big]\gamma ^{0}\gamma ^{5}h^{\mu \nu }D_{\nu }\Psi (x)\Big)\label{Psi}
\end{eqnarray}}%
Here we denote $D_{\mu }=\partial _{\mu }+\frac{1}{8}\omega _{ab\mu }[\gamma
^{a},\gamma ^{b}]$ and $h^{\mu \nu }=g^{\mu \nu }-n^{\mu }n^{\nu }.$ The
4-vector $n$ in the PG reference frame marks the time direction:

\[
n_{\mu }=\left( 1,0,0,0\right) .
\]

Correspondingly $n^{\mu }=\left( 1,-v,0,0\right) .$
The spin connection is given by
\[
\omega _{ab\mu }=\frac{1}{2}(c_{abc}-c_{cab}+c_{bca})E_{\mu }^{c}
\]%
Here $c_{abc}=\eta _{ad}E_{b}^{\mu }E_{c}^{\nu }\partial _{\lbrack \nu
}e_{\mu ]}^{d}$; $g_{\mu \nu }=e_{\mu }^{a}e_{\nu }^{b}\eta _{ab}$. For the
details see Ref. \cite{Alexandrov}, and Refs. \cite%
{Diakonov,Diakonov1,Diakonov2}.
{By $n$ we denote the vector field. Its action may be taken in the form:
 $$
 S_n = \int d^4x \,{\rm det}^{-1}({\bf E}) \,\Big(D_\mu n_\nu D^\mu n^\nu - \lambda (n^2 -1)^2 \Big)
 $$
 At sufficiently large values of $\lambda$ this vector field becomes non - propagating. Its vacuum average gives rise to spontaneous symmetry breaking. }
{The $4$ - vector $n$ in the Painleve - Gullstrand reference frame is constant and marks the time direction:
$$
n_\mu=(1,0,0,0),\,
$$
Correspondingly, $n^\mu = (1,-v,0,0)$. The appearance of this vector breaks the group of general coordinate transformations in four - dimensional space - time to the group of general coordinate transformations in three - dimensional space. As a result three  massless Goldstone modes appear corresponding to the broken boosts. We do not discuss here the physics of these massless excitations. However, we do not exclude that such excitations may play a certain role in the formation of dark matter. }

The only component of the spin connection (in spherical coordinates) is $%
\omega _{r0r}=-\omega _{0rr}=-v^{\prime }(r)$. This term disappears from the
first term in the Painlev\'{e} - Gullstrand reference frame. But it is
essential for the calculation of the stress - energy tensor.

The first step towards the classical theory is the consideration of the
above model with spin neglected. This corresponds to the consideration of
the scalar field instead of the Dirac spinor and the corresponding action.
Hence the theory is that of a scalar field $\Phi $ with the action:
\begin{eqnarray}
S &=&\frac{1}{2}\int d^{4}x\,\sqrt{-g}\,\bar{\Phi}(x)\Big(\frac{1}{\sqrt{-g}}%
\partial _{\mu }\sqrt{-g}g^{\mu \nu }\partial _{\nu }  \nonumber \\
&&+m^{2}+\Big(\frac{1}{\mu \sqrt{-g}}{\partial _{\mu }\sqrt{-g}h^{\mu \nu
}\partial _{\nu }}\Big)^{2}\Big)\Phi (x)  \label{Phi}
\end{eqnarray}%
\rr{There is no precise correspondence between Eqs. (\ref{Psi}) and (\ref{Phi}). In the transition the spin degrees of freedom and the corresponding terms in the lagrangian were  neglected.}
Variation of this action with respect to $\Phi $ gives the wave equation
\[
\Big(\frac{1}{\sqrt{-g}}\partial _{\mu }\sqrt{-g}g^{\mu \nu }\partial _{\nu
}+m^{2}+\Big(\frac{1}{\mu \sqrt{-g}}{\partial _{\mu }\sqrt{-g}h^{\mu \nu
}\partial _{\nu }}\Big)^{2}\Big)\Phi (x)=0
\]%
In the semiclassical limit we substitute $-i\partial _{k}$ by momentum $%
\mathbf{p}$ and $i\partial _{0}$ by energy $\mathcal{E}$. This gives the
following relation for $p_{\mu }=(\mathcal{E},-\mathbf{p})$:
\begin{equation}
\Big(p_{\mu }g^{\mu \nu }p_{\nu }-m^{2}-\Big(\frac{1}{\mu }{p_{\mu }{h^{\mu
\nu }}p_{\nu }}\Big)^{2}\Big)=0  \label{pmu spectrum}
\end{equation}%
In the Painlev\'{e} - Gullstrand reference frame we get:
\begin{equation}
\mathcal{E}^{2}-2\mathcal{E}\mathbf{v}\mathbf{p}+(\mathbf{v}\mathbf{p})^{2}-%
\mathbf{p}^{2}-m^{2}-\frac{\mathbf{p}^{4}}{\mu ^{2}}=0  \label{E spectrum}
\end{equation}%
This equation gives rise to the classical particle Hamiltonian of \cite%
{VolovikBH} given by Eq. (\ref{Energy0}).

\section{The stress - energy tensor of the non - interacting classical
particles}

\label{SectT}

\subsection{General expression for the stress - energy tensor}

In this section we consider matter that consists of the non - interacting
particles. Our aim is to consider the gravitational collapse. Therefore,
following \cite{Kanai:2010ae} we consider the generalization of the Painlev%
\'{e} - Gullstrand spacetime:
\begin{equation}
ds^{2}=dt^{2}-(d\mathbf{r}-\mathbf{v}(\mathbf{r})dt)^{2},  \label{PGQ}
\end{equation}%
with
\begin{equation}
\mathbf{v}=-\frac{1}{m_P }\frac{\mathbf{r}}{r}\,\sqrt{\frac{2m(t,r)}{r}}=-%
\frac{\mathbf{r}}{r}v(r,t)
\end{equation}%
The function $m(t,r)$ is to be obtained through the solution of Einstein
equations.

Although we are going to calculate the stress - energy tensor of the
classical system, we prefer to start from the model of the scalar field with
action Eq. (\ref{Phi}). We will calculate the stress energy tensor of the
corresponding quantum system and take the classical limit at the end of the
calculation. We are to calculate the stress energy tensor through the
variation of action with respect to the variation of metric:
\[
\delta S=\frac{1}{2}\int T_{\mu \nu }\delta g^{\mu \nu }\sqrt{-g}d^{4}x
\]%
Notice that $\delta \sqrt{-g}=-\frac{1}{2}g_{\mu \nu }\sqrt{-g}\delta g^{\mu
\nu }$. In a similar way we calculate the current density as
\[
\delta S=-\int J^{\mu }\delta A_{\mu }\sqrt{-g}d^{4}x
\]

In the semiclassical limit the oscillating factors in the radial wave
functions are given by $\Phi (r,t)\sim e^{iS}\sim e^{-i\mathcal{E}%
t+i\int_{r_{0}}^{r}p(\bar{r})d\bar{r}}$ while the electric current may be
identified with the product of the particle density $\rho $ and the velocity
of substance. This leads to the following semiclassical relation in the
generalized Painlev\'{e} - Gullstrand reference frame:
\begin{eqnarray}
\rho (x,t) &\approx &\Phi ^{\ast }(x,t)\Big(\mathcal{E}(x,t)-\mathbf{v}%
(x,t)\cdot \mathbf{p}(x,t)\Big)\Phi (x,t))  \nonumber \\
\rho (x,t)\mathbf{V}(x,t) &\approx &\Phi ^{\ast }(x,t)\Big(\left( \mathcal{E}%
(x,t)-\mathbf{v}(x,t)\cdot \mathbf{p}(x,t)\right) \mathbf{v}(x,t)+\mathbf{p}%
(x,t)\left( 1+2\frac{|\mathbf{p}(x,t)|^{2}}{\mu ^{2}}\right) \Big)\Phi (x,t)
\end{eqnarray}%
Here $\rho (x,t)$ is the particle density. The number of particles in a
small volume $\Omega $ around the given space - time point is equal to $%
\int_{\Omega }d^{3}x\rho (x,t)$. It is supposed that all these particles
have equal values of energy $\mathcal{E}(x,t)$, velocity $\mathbf{V}(x,t)=%
\hat{\mathbf{r}}V(x,t)$, and momentum $\mathbf{p}(x,t)=\hat{\mathbf{r}}p(x,t)
$. Those quantities are related to each other via the classical equations of
motion
\begin{eqnarray}
V(x,t) &=&\frac{p(x,t)+2p^{3}(x,t)/\mu ^{2}}{\sqrt{m^{2}+{p}^{2}(x,t)+\frac{{%
p}^{4}(x,t)}{\mu ^{2}}}}-v(x,t)  \nonumber \\
\frac{dp(x,t)}{dt} &=&p(x,t)\frac{dv(x,t)}{dr}  \label{Classical}
\end{eqnarray}%
The above equations enable us to relate the absolute value of $\Phi $ with
the physical quantities $\rho $ and $\mathbf{V}$. In an arbitrary reference
frame we have the similar relation
\begin{eqnarray}
j^{\mu }=\frac{1}{\sqrt{-g}}\rho (x,t)\frac{dx^{\mu }}{dt} &\sim &\Phi
^{\ast }(x)\Big(g^{\mu \nu }p_{\nu }-\Big(\frac{2}{\mu ^{2}}{p_{\mu }h^{\mu
\nu }p_{\nu }}\Big)h^{\mu \nu }p_{\nu }\Big)\Phi (x)  \nonumber \\
&=&g^{\mu \nu }|\Phi (x)|^{2}\Big(p_{\nu }(1-2(p^{2}-(pn)^{2})/\mu
^{2})+2n_{\nu }(p^{2}-(pn)^{2})(pn)/\mu ^{2}\Big)
\end{eqnarray}%
Here
\[
\rho (x,t)=\sum_{a}\delta (x-x_{a}(t))
\]%
is the density of particles. Therefore, in the Painlev\'{e} - Gullstrand
coordinates we identify
\[
|\Phi (x,t)|^{2}=\frac{\rho (x,t)}{p^{0}(x,t)}
\]%
while in the arbitrary reference frame
\[
|\Phi |^{2}=\frac{(jp)}{p^{2}-2\frac{(p^{2}-(pn)^{2})^{2}}{\mu ^{2}}}=\frac{%
(jp)}{m^{2}-\frac{(p^{2}-(pn)^{2})^{2}}{\mu ^{2}}}
\]

In the same limit the stress - energy tensor is given by
\begin{eqnarray}
T_{\mu \nu } &\sim &\Phi ^{\ast }(x)(p_{\mu }p_{\nu }-\frac{g_{\mu \nu }}{2}%
(p^{2}-m^{2}))\Phi (x)  \nonumber \\
&&+\Phi ^{\ast }(x)\frac{1}{\mu ^{2}}\Bigl(\frac{g_{\mu \nu }}{2}\left(
p_{\rho }p_{\sigma }h^{\rho \sigma }\right) ^{2}-2\left[ p^{2}-\left(
pn\right) ^{2}\right] p_{\mu }p_{\nu }\Bigr)\Phi (x)
\end{eqnarray}

Equations of motion give
\begin{eqnarray}
T_{\mu \nu } &\sim &\Phi ^{\ast }(x)\Big(p_{\mu }p_{\nu }\left\{ 1-\frac{2}{%
\mu ^{2}}\left[ p^{2}-\left( pn\right) ^{2}\right] \right\} \Bigr)\Phi (x)
\nonumber \\
&=&\frac{p\cdot j}{p^{2}-\frac{2}{\mu ^{2}}\left( {p}^{2}-\left( pn\right)
^{2}\right) ^{2}}p_{\mu }p_{\nu }\left\{ 1-\frac{2}{\mu ^{2}}\left[
p^{2}-\left( pn\right) ^{2}\right] \right\}\label{Tmunu}
\end{eqnarray}

\subsection{The stress energy tensor in the limit $\protect\mu \rightarrow
\infty $}

Let us demonstrate, how Eq. (\ref{Tmunu}) gives rise to the conventional
stress energy tensor of the non - interacting particles in the limit $\mu
\rightarrow \infty $. We have
\[
T^{\mu \nu }=\frac{(jp)}{p^{2}}p^{\mu }p^{\nu }=\frac{\rho }{p^{0}}p^{\mu
}p^{\nu }
\]%
In the generalized PG reference frame we have
\[
j^{\mu }=(\rho ,\rho \mathbf{V}),\,p^{0}=\mathcal{E}-\mathbf{v}\mathbf{p}=%
\frac{m}{\sqrt{1-(\mathbf{V}-\mathbf{v})^{2}}},\,\{-p_{k}\}=\mathbf{p}=\frac{%
m(\mathbf{V}-\mathbf{v})}{\sqrt{1-(\mathbf{V}-\mathbf{v})^{2}}}
\]%
\[
\{p^{k}\}=\frac{m\mathbf{V}}{\sqrt{1-(\mathbf{V}-\mathbf{v})^{2}}}%
,\,(jp)=\rho m\sqrt{1-(\mathbf{V}-\mathbf{v})^{2}}
\]%
This gives
\begin{eqnarray}
T^{00} &=&\rho p^{0}=\rho \frac{m}{\sqrt{1-(\mathbf{V}-\mathbf{v})^{2}}}%
=\rho m\frac{dt}{ds}=\rho m\frac{ds}{dt}u^{0}u^{0}  \nonumber \\
T^{0k} &=&\rho p^{k}=\rho \frac{m{V}^{k}}{\sqrt{1-(\mathbf{V}-\mathbf{v})^{2}%
}}=\rho m\frac{ds}{dt}u^{0}u^{k}  \nonumber \\
T^{jk} &=&\frac{\rho }{p^{0}}p^{j}p^{k}=\rho \frac{mV^{j}{V}^{k}}{\sqrt{1-(%
\mathbf{V}-\mathbf{v})^{2}}}=\rho m\frac{ds}{dt}u^{j}u^{k}
\end{eqnarray}%
Here $u^{j}$ is the four - velocity of the particles/substance. We come to
the conventional expression
\begin{equation}
T_{\mu \nu }=\epsilon u_{\mu }u_{\nu }
\end{equation}
where $\epsilon =\rho m\frac{ds}{dt}$ is the energy density in the rest
frame of medium (notice that $\rho $ is the particle density in the given
reference frame while $\rho \frac{ds}{dt}$ is the particle density in the
rest frame).

\subsection{Expression for the stress - energy tensor for finite $\mu,$ in
the case when the substance is co - moving with the space flow}

In the general case the following expression for the stress energy tensor
should be obtained:
\begin{eqnarray}
T_{\mu\nu} &=& u_{\mu} u_{\nu} f_u(u,n,\mu,\epsilon) + n_\mu n_\nu
f_n(u,n,\mu,\epsilon) + (u_\mu n_\nu + u_\nu n_\mu)f_{un}(u,n,\mu,\epsilon)
+ g_{\mu\nu} f_g(u,n,\mu,\epsilon)
\end{eqnarray}
{Scalar functions $f_u,f_n,f_{un},f_g$ depend on the four
- velocity $u$, the four - vector $n$, constant $\mu$, and the energy
density $\epsilon = \rho m ds/dt$.} In the important particular case, when
velocity of substance $\mathbf{V}$ everywhere is equal to $\mathbf{v}$ the
classical equations of motion give $\mathbf{p} = 0$ and we obtain the
especially simple result:
\begin{eqnarray}
T_{\mu\nu} &=& \epsilon u_{\mu} u_{\nu}\nonumber
\end{eqnarray}

\section{Description of the gravitational collapse in the generalized Painlev%
\'{e} - Gullstrand coordinates}

\label{collapse}

In this section we consider the gravitational collapse of matter that
consists of the non - interacting particles. Those particles being placed
into the Painlev\'{e} - Gullstrand spacetime have the Hamiltonian
\begin{equation}
H(\mathbf{p})=\pm \sqrt{m^{2}+\mathbf{p}^{2}+\frac{\mathbf{p}^{4}}{\mu ^{2}}}%
+\mathbf{p}\,\cdot \mathbf{v}(r,t)  \label{Hamiltonian1}
\end{equation}%
The generalization of the PG spacetime \cite{Kanai:2010ae} has the following
metric
\begin{equation}
ds^{2}=dt^{2}-(d\mathbf{r}-\mathbf{v}(r,t)dt)^{2},  \label{PGQ}
\end{equation}%
with
\begin{equation}
\mathbf{v}=-\frac{1}{m_P }\frac{\mathbf{r}}{r}\,\sqrt{\frac{2m(t,r)}{r}}%
=-v\left( r,t\right) \mathbf{\hat{r}}
\end{equation}%
The function $m(t,r)$ is to be obtained through the solution of Einstein
equation.

It was shown above that the noninteracting matter with the Hamiltonian of
\cite{VolovikBH} has the stress energy tensor equal to that of the
conventional matter in the generalized Painlev\'{e} - Gullstrand coordinates
in the important particular case, when at each point the velocity of matter
is precisely equal to $\mathbf{v}$. The problem for the gravitational
collapse of matter with this stress - energy tensor is solved in \cite%
{Kanai:2010ae}. We repeat here the solution for completeness.

In the spherical coordinates the Einstein equation $8\pi T^{\mu }{}_{\nu
}=R^{\mu }{}_{\nu }-(1/2)\delta ^{\mu }{}_{\nu }R$ receives the form (we use
in this section the system of units with $m_P =1$):
\begin{eqnarray}
8\pi T^{0}{}_{0} &=&-\frac{2m^{\prime }}{r^{2}},  \label{Eeq-00} \\
8\pi T^{1}{}_{0} &=&\frac{2\dot{m}}{r^{2}},  \label{Eeq-10} \\
8\pi T^{1}{}_{1} &=&-\frac{2m^{\prime }}{r^{2}}+\frac{2\dot{m}}{{r^{2}}}%
\left( \frac{2m}{r}\right) ^{-1/2},  \label{Eeq-11} \\
8\pi T^{2}{}_{2} &=&8\pi T^{3}{}_{3}=-\frac{m^{\prime \prime }}{r}+\left(
\frac{\dot{m}}{2r^{2}}+\frac{\dot{m}^{\prime }}{r}\right) \left( \frac{2m}{r}%
\right) ^{-1/2}-\frac{\dot{m}m^{\prime }}{r^{2}}\left( \frac{2m}{r}\right)
^{-3/2},  \label{Eeq-22}
\end{eqnarray}%
Here dots represent the differentiation with respect to time while $^{\prime
}$ is the differentiation with respect to the radial coordinate $r$. From
these equations one has the identity
\begin{equation}
T^{1}{}_{1}=T^{0}{}_{0}+T^{1}{}_{0}\left( \frac{2m}{r}\right) ^{-1/2}.
\label{T11}
\end{equation}

For the noninteracting particles (perfect fluid) in the above particular
case the energy - momentum tensor $T^{\mu }{}_{\nu }$ has the form:
\begin{equation}
T^{\mu \nu }=\epsilon u^{\mu }u^{\nu }.  \label{T}
\end{equation}%
Here $\epsilon $ is the energy density, and $u^{\mu }=\left(
1,\,-v(t,r),\,0,\,0\right) $ the radial four-velocity of the fluid.
Correspondingly $u_{\mu }=\left( 1,\,0,\,0,\,0\right) $. For the
definiteness let us assume that $\mathbf{v}$ is directed along the $x$ axis.
Then
\begin{equation}
g^{\mu \nu }=\left(
\begin{array}{cccc}
1 & -{v} & 0 & 0 \\
-{v} & v^{2}-1 & 0 & 0 \\
0 & 0 & -1 & 0 \\
0 & 0 & 0 & -1%
\end{array}%
\right) ,\quad g_{\mu \nu }=\left(
\begin{array}{cc}
1-v^{2} & \mathbf{v} \\
\mathbf{v}^{T} & -1%
\end{array}%
\right)
\end{equation}

It follows from Eqs. (\ref{T}) and (\ref{T11}) that $v(t,r)$ is given by
\begin{equation}
v(t,r)=\sqrt{\frac{2m(t,r)}{r}}.  \label{v}
\end{equation}

Next, the integration of the above mentioned equations of motion gives
\begin{equation}
m(t,r)=4\pi \int_{0}^{r}\epsilon (t,r)r^{2}dr  \label{m}
\end{equation}%
and
\begin{equation}
\epsilon =\frac{1}{6\pi t^{2}},\,t<0.  \label{e}
\end{equation}

We come to the following pattern of the gravitational collapse. If the
velocity of matter at the starting moment $t_{0}$ is equal to the function $%
v(r,t_{0})$ of the generalized Painlev\'{e} - Gullstrand reference frame,
and everywhere the three - momentum of the particles vanishes at $t=t_{0}$,
then the Einstein equations have the solution given by Eqs. (\ref{v}), (\ref%
{m}), (\ref{e}). Thus, matter contracts towards the center of the BH
together with the "falling" space. This gives
\[
m(t,r)=\frac{2r^{3}}{9t^{2}}m_P^{2},\,v(t,r)=\frac{2r}{3|t|}
\]%
(\rrv{We restore here the expression that contains the Planck mass $m_P$ explicitly due to unit considerations}.) As a
result the position of the horizon (where the velocity equals $1$) depends
on time:
\[
r_{h}=3|t|/2
\]%
One can see, that at each finite value of $t$ the velocity of space fall
vanishes for $r=0$ only. The space - time metric remains regular everywhere.

The collapse of dust placed within a sphere leads at $t\rightarrow -0$ to
the formation of the ordinary Painlev\'{e} - Gullstrand BH (see \cite%
{Kanai:2010ae}). In the next section we will consider the classical motion
of particles in the formed BH regularized in the small vicinity of its
center as explained in Sect. \ref{Section_DF}.

\section{Classical dynamics of particles}

\label{SectClass}

\begin{figure}[h]
\centering
\includegraphics[width=10cm]{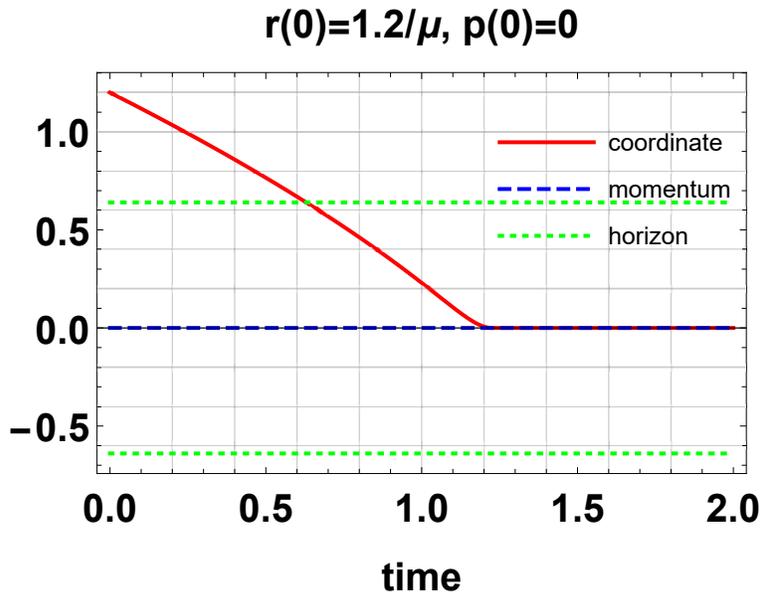}
\caption{The radial trajectory of the particle that falls down to the black
hole (red solid line): the dependence of radial coordinate in the units of $%
1/m_P$ on time (in the same units); radial momentum in the units of $m_P$ as
a function of time (dashed blue line). The values of parameters $M = 0.5 \,
m_P$, $m=0.1 \, m_P$, $\mu = m_P$, while $Q = 0.4$. The
external horizon is represented by the dotted green line. The particle
starts falling at $r(0)=1.2/\mu$ and $p(0) = 0$. One can see, that this
particle falls together with "vacuum". It reaches the center of the BH and
stays there.}
\label{fig.t}
\end{figure}

The classical Hamiltonian of the quasiparticles in the presence of the black
hole has the form:
\begin{equation}
H(\mathbf{p})=\pm \sqrt{m^{2}+\mathbf{p}^{2}+\frac{\mathbf{p}^{4}}{\mu^{2}}%
}-\mathbf{p}\hat{\mathbf{r}}\,v(r)  \label{Hamiltonian1}
\end{equation}%
For the radial motion the corresponding classical equations are:
\begin{eqnarray}
\frac{dr}{dt} &=&\frac{p+2p^{3}/\mu^{2}}{\sqrt{m^{2}+{p}^{2}+\frac{{p}^{4}%
}{\mu^{2}}}}-v(r)  \nonumber \\
\frac{dp}{dt} &=&p\frac{dv(r)}{dr}
\end{eqnarray}%
The Fermi surface for the particular value of $r$ crosses the axis of radial
momentum at
\[
p_{\pm }/\mu=\sqrt{\frac{v^{2}(r)-1}{2}\pm \sqrt{\frac{(v^{2}(r)-1)^{2}}{4}%
-\frac{m^{2}}{\mu}}}
\]%
One can see, that the Fermi surface is not formed immediately behind the
external horizon. Instead, it is formed at
\[
v(r)\geq \sqrt{1+2m/\mu}
\]%
\begin{figure}[h]
\centering
\includegraphics[width=10cm]{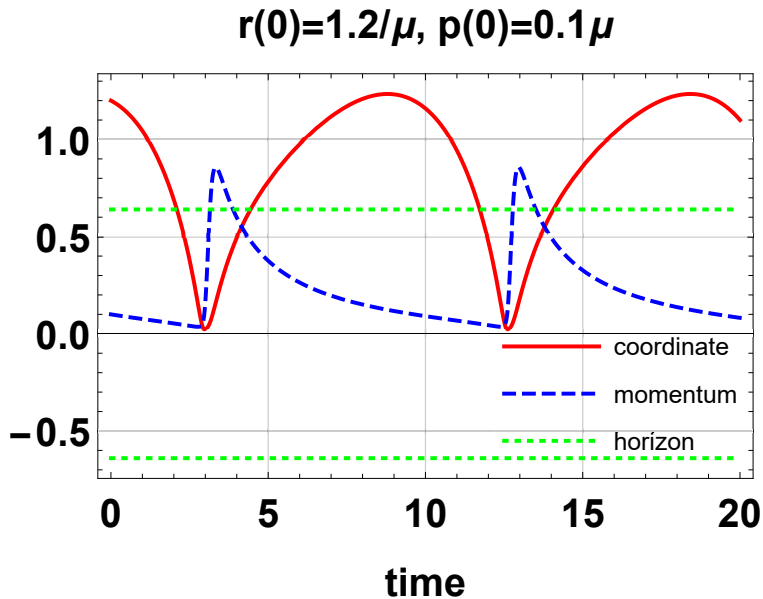}
\caption{The radial trajectory of the particle that falls down to the black
hole and escapes from it (red solid line): the dependence of radial
coordinate in the units of $1/m_P$ on time (in the same units); radial
momentum in the units of $m_P$ as a function of time (dashed blue line).
The values of parameters $M=0.5\,m_P$, $m=0.1\,m_P$, $\mu = m_P$, $Q=0.4$. The external horizon is represented by the
dotted green line. The particle starts falling at $r(0)=1.2/\mu$ and $%
p(0)=0.1\mu$. This particle falls more slow than "vacuum". It reaches the
vicinity of the center of the BH. There the repulsion force pushes it away,
it escapes from the BH, then falls again, etc.}
\label{fig.t1}
\end{figure}

\begin{figure}[h]
\centering
\includegraphics[width=10cm]{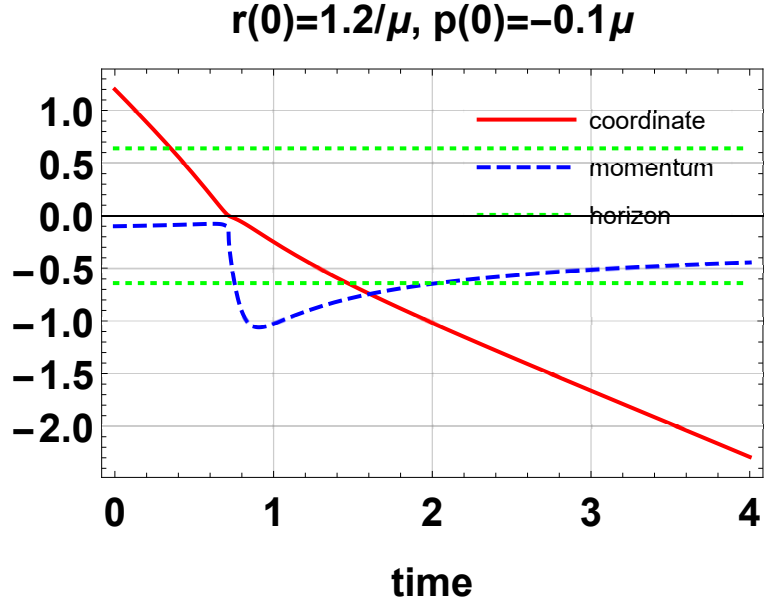}
\caption{The radial trajectory of the particle that traverses the black hole
(red solid line): the dependence of radial coordinate in the units of $1/m_P$
on time (in the same units); radial momentum in the units of $m_P$ as a
function of time (dashed blue line). The values of parameters $M = 0.5 \,
m_P $, $m=0.1 \, m_P$, $\mu = m_P$, $Q = 0.4$. The
external horizon is represented by the dotted green line. The particle
starts falling at $r(0)=1.2/\mu$ and $p(0) = -0.1 \mu$. This particle falls
faster than "vacuum". It reaches the center of the BH, crosses it. The
repulsion force accelerates it, and the particle escapes at the diametrically opposite point.}
\label{fig.tm1}
\end{figure}

The typical classical trajectories of the particles are calculated and
represented in Figs. \ref{fig.t}, \ref{fig.t1}, \ref{fig.tm1}. The particles
that fall together with the "vacuum" reach the center of the black hole and
stay there. However, if the initial momentum is nonzero, the particles that
have fallen down to the black hole receive large values of radial momentum
in a small vicinity of the BH center. As a result they escape from the BH.
If the initial momentum was directed to the center of the BH, then the
particle traverses the BH and escapes it from a diametrically opposite
point. If the initial momentum is in the opposite direction, then its
velocity reverses the sign close to the center of the BH, the particle is
turned back. In the exterior of the black hole the momentum is decreased,
and the particle velocity changes the sign again. The particle falls down
again thus forming oscillations.

\begin{figure}[h]
\centering
\includegraphics[width=10cm]{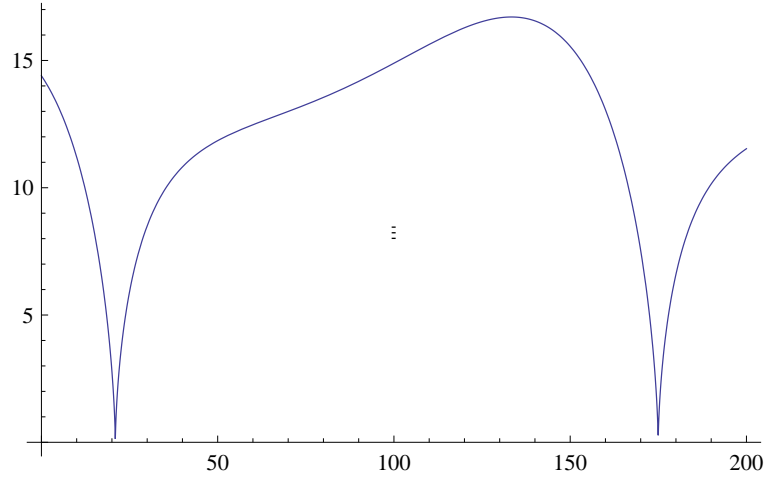}
\caption{The radial trajectory of the particle that falls down to the black
hole and escapes from it: the dependence of radial coordinate in the units
of $1/m_P$ on time. The values of parameters $M = 6 \, m_P$, $m=0.01 \, m_P$, $\mu = m_P$, $Q = 0.1$. The external horizon $h$ is
not represented here, but the motion starts at $r(0)=1.2\,h$ and $p(0) =
0.01 \mu$. It is supposed, that this figure represents qualitatively the
typical trajectory for $M \gg \mu \sim m_P \gg m$.}
\label{fig.100}
\end{figure}

Our numerical data allow to estimate the typical time period of those
oscillation for $m\ll \mu \sim m_P \ll M$ (when the amplitude is of the order of the
horizon) $T \sim 20 \, M/\mu^2$ (see Fig. \ref{fig.100}). This value for the
solar mass BH (in seconds) is
\[
T_\odot \sim 20\, \frac{2\, \cdot 10^{30} Kg}{4\,\cdot 10^{-16} Kg^2} 
\frac{\hbar}{c^2}\sim  10^{-4} s
\]

\section{Conclusions}

\label{SectConcl}

To conclude, in the present paper we consider the model of noninteracting
Dirac fermions with the modified Hamiltonian proposed in \cite{VolovikBH}.
In this model the extra $\sim \mathbf{p}^{2}$ term is added to the Dirac
Hamiltonian. First of all, we propose the covariant generalization of this
model. The resulting field theory is defined in terms of the Dirac spinor
field. It depends on the background field $n_{\nu }$ that equals $n_{\mu }=$
$(1,0,0,0)$ in the Painlev\'{e} - Gullstrand reference frame. The field $n$
points into the direction of time in this coordinate system. It appears as a result of the spontaneous
symmetry breaking. This  symmetry breaking also leads to the appearance of the massless Goldstone modes. \rr{The direct interaction term of those modes with matter is suppressed by the factor $1/\mu$, where $\mu$ is of the order of Plank mass.} But
they are coupled to gravity. The consideration of the physics of those modes
is out of the scope of the present paper. But we do not exclude, that, in
certain theoretical schemes, they contribute the dark matter.

Next, we consider the classical limit of the obtained system. First,
neglecting spin we come to the theory of the scalar field with a certain
action (depending on $n$). Next, we consider the semiclassical approximation
within this theory, which gives both the classical particle Hamiltonian of
\cite{VolovikBH}, and the implicitly defined expression for the stress -
energy tensor of medium that consists of the noninteracting particles. It
appears, that the Einstein equations admit the gravitational collapse
solution identical to that of the system of the conventional noninteracting
particles. This solution describes the dust falling together with space -
time.

Finally we describe the dynamics of particles in the presence of the
existing black holes. It appears, that only those particles remain inside
the BH, which fall with velocity $v$ entering the expression for the Painlev%
\'{e} Gullstrand metric. The particles that fall towards the center of the
BH with nonzero momentum either pass through the BH and reach infinity or
turn back at the small vicinity of the BH center, escape from the BH, and
proceed the oscillatory motion. Sure, this means that the particles are able
to move with the velocity larger than the speed of light. Although the
considered theory is manifestly covariant (as explained in Sect. \ref%
{SectCov}), the geodesic lines are already not the solutions of the
classical equations of motion of point-like particles. The solutions of
those equations may correspond to the space - like pieces of the particle
worldlines, which does not contradict the general covariance.

According to our estimates for the BH with the solar mass the typical period
of the mentioned oscillations (when interactions are neglected) is smaller
than one second. This means, that if the effective Hamiltonian of particles
indeed receives the considered contribution from Planck physics, then we
cannot ignore it in the dynamics: matter that has fallen to the BH escapes
from it within the observable period of time.


One of the authors (M.A.Z.) kindly acknowledges useful discussions with G.E. Volovik.

\vspace{6pt}











\end{document}